\PassOptionsToPackage{usenames,dvipsnames}{xcolor}
\documentclass[arXiv]{actapoly}
 
\usepackage[english]{babel}
\usepackage[utf8]{inputenc}
\usepackage{amsfonts}
\usepackage{amsmath}
\usepackage{amssymb}
\usepackage{amsthm}
\usepackage{interval}
\usepackage{bm}
\usepackage{soul}
\usepackage[normalem]{ulem}
\usepackage{xifthen}
\usepackage{scalerel}
\usepackage{multicol}
\usepackage{multirow}
\usepackage{csquotes}
\usepackage{subcaption}
\usepackage[ruled,vlined,linesnumbered]{algorithm2e}
\usepackage{xcolor}

\def\HiLiY{\leavevmode\rlap{\hbox to \hsize{\color{yellow!50}\leaders\hrule height .8\baselineskip depth .5ex\hfill}}}
\def\HiLiG{\leavevmode\rlap{\hbox to \hsize{\color{green!50}\leaders\hrule height .8\baselineskip depth .5ex\hfill}}}

\makeatletter
\let\r@FirstPage\relax
\let\r@LastPage\relax
\makeatother

\soulregister\cite7
\soulregister\ref7
\soulregister\pageref7

\DeclareGraphicsExtensions{.pdf,.eps,.png,.jpg,.svg}	
\graphicspath{ {./figures/Final} }

\newcommand{\Fref}[1]{Fig.~\ref{#1}}
	
\newcommand{\Eref}[1]{Eq.~(\ref{#1})}

\newcommand{\Sref}[1]{{Section~\ref{#1}}}

\newcommand{\tiledom}[1]{\ensuremath{\mathcal{T}\ifthenelse{\isempty{#1}}{}{_{(#1)}} }}
\newcommand{\conog}[1]{\ensuremath{\mathcal{G}\ifthenelse{\isempty{#1}}{}{_{(#1)}} }}		

\newcommand{\bmath}[1]{\ensuremath{\bm{#1}}}

\mathchardef\mhyphen="2D

\newcommand{\atx}{\ensuremath{(\tens{x})}}

\newcommand{\set}[1]{{\mathbb #1}}

\newcommand{\setRn}[1]{\set{R}^{#1}}

\newcommand{\domain}{\Omega}

\newcommand{\tens}[1]{\bm{#1}}				            
\newcommand{\tensf}[1]{\bmath{\mathbf{#1}}} 			

\newcommand{\trn}{^{\sf T}}
\newcommand{\invtrn}{^{-{\sf T}}}

\newcommand{\mtrx}[1]{\begin{bmatrix}#1\end{bmatrix}}

\newcommand{\semtrx}[1]{\mathsf{#1}} 					
\newcommand{\sevek}[1]{\mathsf{#1}} 					

\begin{document}

\title[FETI methods for topology optimization problems]{Comparison of FETI-based domain decomposition methods for topology optimization problems}

\correspondingauthor[T. Medřický]{Tomáš Medřický}{k132}{tomas.medricky@fsv.cvut.cz}
\author[M. Doškář]{Martin Doškář}{k132}
\author[I. Pultarová]{Ivana Pultarová}{k101}
\author[J. Zeman]{Jan Zeman}{k132}

\institution{k132}{Czech Technical University in Prague, Faculty of Civil Engineering, Department of Mechanics, Thákurova 7, 166 29 Prague, Czech Republic}
\institution{k101}{Czech Technical University in Prague, Faculty of Civil Engineering, Department of Mathematics, Thákurova 7, 166 29 Prague, Czech Republic}

\begin{abstract}
We critically assess the performance of several variants of dual and dual-primal domain decomposition strategies in problems with fixed subdomain partitioning and high heterogeneity in stiffness coefficients typically arising in topology optimization of modular structures.
Our study considers Total FETI and FETI Dual-Primal methods along with three enhancements: $k$-scaling, full orthogonalization of the search directions, and considering multiple search-direction at once, which gives us twelve variants in total.
We test these variants both on academic examples and snapshots of topology optimization iterations.
Based on the results, we conclude that (i)~the original methods exhibit very slow convergence in the presence of severe heterogeneity in stiffness coefficients, which makes them practically useless, (ii) the full orthogonalization enhancement helps only for mild heterogeneity, and (iii) the only robust method is FETI Dual-Primal with multiple search direction and $k$-scaling.
\end{abstract}


\maketitle

\section{Introduction}
Topology optimization has became an indispensable tool in the design process, allowing for optimal distribution of a material within a provided space with respect to given criteria (such as minimization of compliance, maximization of output response of mechanisms, or increasing the lowest eigenfrequency) under given constraints (typically limiting the available material volume).
In Solid Isotropic Material with Penalization (SIMP) method~\cite{bendsoe_topology_2004}---one of the most common approaches to topology optimization---the material distribution is parameterized with a scalar relative density field $\rho\atx$, with $0 \leq \rho \leq 1$, which consequently governs the distribution of stiffness parameters in a domain such that
\begin{equation}
    \tensf{E}\atx = \tensf{E}_\text{min} + \rho^{p}\atx \left( \tensf{E}_{0} - \tensf{E}_{\text{min}} \right) \,,
    \label{eq:SIMPstiffness}
\end{equation}
where $p$ is a given penalization coefficient (usually $p\geq3$), helping the optimization procedure to achieve the desired \enquote{0-1} design without regions of intermediate densities, and $\tensf{E}_{0}$ and $\tensf{E}_{\text{min}}$ are stiffness tensors of the solid material and voids, respectively. 
Note that while the stiffness tensor of the bulk material is physical, void stiffness serves only numerical purposes of avoiding an indefinite Hessian matrix. On one hand, it should be small enough in order to model voids properly, on the other hand too small values lead to ill-posed state problem.
The optimal design is then usually sought-for in an iterative manner, alternately solving the state equations with a fixed densities and updating design density variables according to the sensitivities computed for a current design.

Being a matured technology, topology optimization is now widely used in industrial applications with millions of unknowns, and first applications reaching billions of unknowns are emerging~\cite{aage_giga-voxel_2017}. 
Consequently, such scales of optimization problems often prohibit the use of direct solvers in favour of iterative ones, which are further combined with parallelisation. 
Domain decomposition (DD) techniques thus seem promising candidates; however, the presence of high contrast in coefficients due to the stiffness parameterization, recall~\Eref{eq:SIMPstiffness}, may cause slower convergence and consequently poor performance of iterative solvers based on DD, as has been already shown in~\cite{arul_feti_2020}.

To make the task even more challenging, in this contribution, we focus in particular on modular topology optimization problems, recently introduced in~\cite{bib:TrussStructures,tyburec_modular_2022}, in which the main domain is composed of a limited number of square modules. The multiple occurrence of individual modules can thus be readily exploited to accelerate calculations; on the other hand, the modularity automatically introduces fixed, regular partitioning, which---as will be shown later---is detrimental for the performance of DD-based solvers.
Admittedly, the modular grid constitutes only the coarsest partitioning and more refined divisions of each module following e.g. the distribution of a material with the module can be performed, but we did not follow this possibility in our current study. 

After recalling the fundamentals of two Finite Element Tearing and Interconnecting methods in~\Sref{sec:FETI_description}, we review several modifications aimed at improving the performance and robustness of the solution strategy; see~\Sref{sec:FETI_improving_variants}. Finally, in~\Sref{sec:Numerical_tests}, the performance of the original methods and their modifications is assessed on two test sets: three academic problems from the literature and three geometries arising from three distinct snapshots of modular topology optimization of the emblematic Messerschmitt–Bölkow–Blohm beam.

\section{Dual and dual-primal domain decomposition methods}
\label{sec:FETI_description}

Characteristically for domain decomposition methods, we assume that the original domain $\domain$ is partitioned into $N_{s}$ mutually disjoint subdomains $\domain^{s}$ with $s = 1\dots N_{s}$. The solution to the system of linear equations 
\begin{equation}
    \semtrx{K}^{\Omega} \sevek{u}^{\Omega} = \sevek{f}^{\Omega} \,,
\end{equation}
which arises from e.g. numerically discretized partial differential equations valid in $\domain$, is then sought for via a series of subdomain-wise problems
\begin{equation}
    \semtrx{K}^{s} \sevek{u}^{s} = \sevek{f}^{s} \quad \text{for} \quad s = 1 \dots N_{s} \,,
    \label{eq:localequilibrium}
\end{equation}
with an additional constraint ensuring continuity of the solution $\left\{\sevek{u}^{s}\right\}_{s=1\dots N_{s}}$ across subdomain boundaries. 
Two major branches of domain decomposition methods can be distinguished based on the manner in which the aforementioned continuity is enforced: primal decomposition methods, such as the Schur Complement method~\cite{kruis_domain_2006}, satisfy the continuity by construction of an approximation space, while the dual methods, e.g. Finite Element Tearing and Interconnecting (FETI) method~\cite{farhat_unconventional_1992}, impose the continuity requirement as an equality constraint enforced via Lagrange multipliers. 
Finally, as its name suggests, the Finite Element Tearing and Interconnecting Dual-Primal (FETI-DP) method~\cite{bib:feti-dp} adopted in our work combines both approaches.

In the following two subsections, we briefly summarize the essentials of both dual and dual-primal methods.

\subsection{Total-FETI}
\label{sec:TFETI}

The Total FETI method (T-FETI), introduced by Dostál and co-workers~\cite{bib:tfeti} and adopted here as a representative of the dual domain decomposition methods, is closely related to the original FETI method~\cite{farhat_unconventional_1992} with the main difference being the way the Dirichlet boundary conditions (BC) are imposed.
In T-FETI, these BC are handled similarly to enforcing the continuity across subdomain boundaries, i.e. the constraints posed on the domain-wise displacements $\sevek{u}^{s}$ read as
\begin{equation}
	\sum_{s=1}^{N_{s}} \semtrx{B}^{s} \sevek{u}^{s} = \semtrx{B} \sevek{u} = \sevek{c} \,
	\label{eq:tfeti_constraint}
\end{equation}
where $\sevek{u}$ collects the domain-wise unknowns such that
\begin{equation}
	\sevek{u} = \mtrx{ {\sevek{u}^{1}}\trn, {\sevek{u}^{2}}\trn, \dots , {\sevek{u}^{N_{s}}}\trn }\trn
	\label{eq:tfeti_all_domain_displacement}
\end{equation} 
and $\semtrx{B}$ combines the individual $\semtrx{B}^{s}$'s accordingly. Thus, in contrast to the standard FETI, constraint~(\ref{eq:tfeti_constraint}) features a right-hand side vector $\sevek{c}$ that is in general nontrivial; in addition to the zero entries pertinent to the cross-boundary continuity, it also contains values related to the Dirichlet BC.
As a result, the number of rigid body modes of each domain is only problem specific (e.g. three for two-dimensional mechanics problems) and factorization can be recycled for all occurrences of one module irrespective of whether the occurrence is supported or floating.

Combining Eqs. (\ref{eq:localequilibrium}) and (\ref{eq:tfeti_constraint}) leads to 
\begin{equation}
	\mtrx{ \semtrx{K} & \semtrx{B}\trn \\ \semtrx{B} & \semtrx{0} }
	\mtrx{ \sevek{u} \\ \sevek{\lambda} }
	=
	\mtrx{ \sevek{f} \\ \sevek{c} }
	\label{eq:tfeti_primaldual}
\end{equation}
where $\semtrx{K}$ is a block-diagonal matrix composed of $\semtrx{K}^{s}$ and $\sevek{f}$ arises from $\sevek{f}^{s}$ concatenated similarly to $\sevek{u}$ in~\Eref{eq:tfeti_all_domain_displacement}.

Assuming the $\sevek{\lambda}$ is known and provided that $\sevek{f} - \semtrx{B}\trn\sevek{\lambda}$ is orthogonal to the nullspace of $\semtrx{K}$, $\sevek{u}$ can be uniquely determined up to rigid body modes of individual domains expressed as $\semtrx{R}\,\sevek{\alpha}$, where $\semtrx{R}$ is a matrix containing the nullspace vectors of each domain and $\sevek{\alpha}$ is a vector of rigid body modes coefficients.
Without diving into technical details, for which we refer an interested reader to e.g. \cite{dostal_total_2006}, the original formulation can be recast into the dual form
\begin{equation}
	\mtrx{ \semtrx{F} & \semtrx{G} \trn \\ \semtrx{G} & \semtrx{0}} 
	\, 
	\mtrx{ \sevek{\lambda} \\ \sevek{\alpha}} 
	= 
	\mtrx{\sevek{d} \\ \sevek{e}}
	\label{eq:tfeti:FGG0=lambdaalfa}
\end{equation}
with
\begin{align*}
	\semtrx{F} &= \semtrx{B} \semtrx{K}^{\dagger} \semtrx{B}\trn \,,		&	\sevek{d} &= \semtrx{B} \semtrx{K}^{\dagger} \sevek{f} - \sevek{c} \,, \\
	\semtrx{G} &= - \semtrx{R}\trn \sevek{B}\trn \,,						&	\sevek{e} &= - \semtrx{R}\trn \sevek{f} \,,
\end{align*}
where $\semtrx{K}^{\dagger}$ denotes the Moore-Penrose pseudo-inverse of $\semtrx{K}$.
Problem~(\ref{eq:tfeti:FGG0=lambdaalfa}) is traditionally solved with a projected preconditioned conjugate gradient method.

\subsection{FETI-DP}
\label{sec:FETIDP}

In Finite Element Tearing and Interconnecting Dual-Primal (FETI-DP) method \cite{bib:feti-dp}, the domain-wise displacement field $\sevek{u}^{s}$ is decomposed into two parts:
\begin{enumerate}
	\item boundary degrees of freedom $\sevek{u}_{c}^{s}$ whose continuity is directly enforced in the primal unknowns, and 
	\item remaining degrees of freedom $\sevek{u}_{r}^{s}$ that contain both internal degrees of freedom, which do not directly communicate with other domains, and boundary degrees of freedom whose continuity is enforced via equality constraints, similarly to \Eref{eq:tfeti_constraint}.
\end{enumerate}
Without loss of generality, we assume that $\sevek{u}^{s}$ is ordered such that 
\begin{equation}
	\sevek{u}^{s} 
	=
	\mtrx{
		\sevek{u}^{s}_{r} \\
		\sevek{u}^{s}_{c} 
	}\!,
	\;
	\semtrx{K}^{s} 
	=
	\mtrx{
		\semtrx{K}^{s}_{rr} & \semtrx{K}^{s}_{rc}\\
		\semtrx{K}^{s}_{cr} & \semtrx{K}^{s}_{cc} 
	}\!,
	\;\;
	\text{and}
	\;\;
	\sevek{f}^{s} 
	=
	\mtrx{ 
		\sevek{f}^{s}_{r} \\
		\sevek{f}^{s}_{c} 
	}\!.
	\label{eq:fetidp:local_split}
\end{equation}
Furthermore, we consider a global vector $\sevek{u}_{c}$ that stores the shared boundary degrees of freedom (DOFs) and a Boolean matrix $\semtrx{B}^{s}_{c}$ such that
\begin{equation}
	\sevek{u}^{s}_{c} = \semtrx{B}^{s}_{c} \sevek{u}_{c} \,.
	\label{eq:fetidp:global_primal}
\end{equation}
The equality constraint ensuring continuity of the remaining $\sevek{u}^{s}_{b}$ then reads 
\begin{equation}
	\sum_{s=1}^{N^{s}} \semtrx{B}^{s}_{r} \sevek{u}^{s}_{r} = \semtrx{B}_{r} \sevek{u}_{r} = \sevek{0} \,,
	\label{eq:fetidp:global_constraint}
\end{equation}
where---analogously to Eqs. (\ref{eq:tfeti_constraint}) and (\ref{eq:tfeti_all_domain_displacement})---$\sevek{u}_{r}$ and $\semtrx{B}_{r}$ are concatenations of $\sevek{u}^{s}_{r}$ and $\semtrx{B}^{s}_{r}$, respectively.

Expressing the equilibrium (\ref{eq:localequilibrium}) and the constraint~(\ref{eq:fetidp:global_constraint}) in terms of the quantities introduced in Eqs. (\ref{eq:fetidp:local_split})--(\ref{eq:fetidp:global_constraint}) leads to
\begin{equation}
	\mtrx{
		\semtrx{K}_{cc} & \semtrx{K}_{cr} & \semtrx{0} \\
		\semtrx{K}_{rc} & \semtrx{K}_{rr} & \semtrx{B}_{r}\trn \\
		\semtrx{0}      & \semtrx{B}_{r}  &  \semtrx{0} 
	}
	\mtrx{
		\sevek{u}_{c} \\
		\sevek{u}_{r} \\
		\sevek{\lambda}
	}
	=
	\mtrx{
		\sevek{f}_{c} \\
		\sevek{f}_{r} \\
		\sevek{0}
	}
	\!,
	\label{eq:fetidp:primal}
\end{equation}
with
\begin{align*}
	\semtrx{K}_{cc} &= \sum_{s=1}^{N_{s}} {\semtrx{B}^{s}_{c}}\trn \semtrx{K}^{s}_{cc} \semtrx{B}^{s}_{c} \,,
	\quad\quad\quad\quad\quad\quad\!
	\sevek{f}_{c} = \sum_{s=1}^{N_{s}} {\semtrx{B}^{s}_{c}}\trn \sevek{f}^{s}_{c} \,,
	\\
	\semtrx{K}_{cr} &= \semtrx{K}_{rc}\trn = \mtrx{ {\semtrx{B}^{1}_{c}}\trn \semtrx{K}^{1}_{cr} &  {\semtrx{B}^{2}_{c}}\trn \semtrx{K}^{2}_{cr} & \dots &  {\semtrx{B}^{N_{s}}_{c}}\trn \semtrx{K}^{N_{s}}_{cr} } \!,
	\\
	\semtrx{K}_{rr} &= \mtrx{ 
		\semtrx{K}^{1}_{rr} & \semtrx{0} 		  & \dots  & \semtrx{0} \\
		\semtrx{0}          & \semtrx{K}^{2}_{rr} & \dots  & \semtrx{0}  \\
		\vdots              & \vdots 			  & \ddots & \vdots \\
		\semtrx{0}          & \semtrx{0} 		  & \dots  & \semtrx{K}^{N_{s}}_{rr}
	} ,
	\;\;
	\text{and}
	\;\;
	\sevek{f}_{r} = \mtrx{ 
		\sevek{f}^{1}_{r} \\
		\sevek{f}^{2}_{r} \\
		\vdots \\
		\sevek{f}^{N_{s}}_{r}
	}\!.
\end{align*}

Finally, due to the domain-wise nature of the second row in \Eref{eq:fetidp:primal}, $\sevek{u}^{s}_{r}$ can be computed in parallel from the known $\sevek{u}_{c}$ and $\sevek{\lambda}$; hence, plugging the expression for $\sevek{u}^{s}_{r}$ in terms of the remaining quantities into the first and the third rows of \Eref{eq:fetidp:primal} leads to the final dual-primal form
\begin{equation}
	\mtrx{
		\semtrx{F}_{rr}     &  \semtrx{F}_{rc} \\
		\semtrx{F}_{rc}\trn & -\bar{\semtrx{K}}_{cc}
	}
	\,
	\mtrx{
		\sevek{\lambda} \\
		\sevek{u}_{c}
	} 
	= 
	\mtrx{
		\sevek{d}_{r} \\
	   -\bar{\sevek{f}}_{c}
	}
	\!,
	\label{eq:fetidp:interface}
\end{equation}
where
\begin{align*}
	\semtrx{F}_{rr} &= \semtrx{B}_{r} \semtrx{K}^{-1}_{rr} \semtrx{B}_{r}\trn \,,		
	&
	\semtrx{F}_{rc} &= \semtrx{B}_{r} \semtrx{K}^{-1}_{rr} \semtrx{K}_{rc} \,,
	\\	
	\bar{\semtrx{K}}_{cc} &= \semtrx{K}_{cc} - \semtrx{K}_{cr} \semtrx{K}_{rr}^{-1} \semtrx{K}_{rc} \,,
	&
	\sevek{d}_{r} &= \semtrx{B}_{r} \semtrx{K}_{rr}^{-1} \sevek{f}_{r} \,,
	\\
	\bar{\sevek{f}}_{c} &= \sevek{f}_{c} - \semtrx{K}_{cr} \semtrx{K}_{kk}^{-1} \sevek{f}_{r} \,.
\end{align*}
Note that DOFs in $\sevek{u}_{c}$ are chosen such that there is no need for a pseudo-inverse of $\semtrx{K}_{rr}$ in contrast to T-FETI. In FETI-DP, DOFs related to corner nodes in a regular subdomain partitioning typically constitute~$\sevek{u}_{c}$.

Problem~(\ref{eq:fetidp:interface}) is traditionally solved iteratively with a preconditioned conjugate gradient method with $\sevek{u}_{c}$ being condensed out, resulting in a problem dependent solely on $\sevek{\lambda}$~\cite{bib:feti-dp}.

\section{Enhancements for heterogeneous domains}
\label{sec:FETI_improving_variants}

\subsection{$k$-scaling}
\label{sec:k-scaling}

The merit of both formulations described above is that they can be conveniently parallelized thanks to their additive structure. In both, the solution is sought with an iterative conjugate gradient method considering only the first block of the corresponding equations (\ref{eq:tfeti:FGG0=lambdaalfa}) and (\ref{eq:fetidp:interface}), while the remaining blocks are incorporated either via projection (T-FETI) or by condensating out the primal unknowns (FETI-DP).

Distinctively for domain decomposition methods, the iterative procedure also benefits from a coarse problem preconditioner $\bar{\semtrx{F}}^{-1}$ of either $\semtrx{F}$, recall \Eref{eq:tfeti:FGG0=lambdaalfa}, or $\semtrx{F}_{rr}$, \Eref{eq:fetidp:interface}, constructed as a sum of local inverses
\begin{equation}
    \bar{\semtrx{F}}^{-1} 
    =
    \sum_{s=1}^{N_{s}} 
    \left( 
    \tilde{\semtrx{B}}^{s} 
    \mtrx{0 & 0 \\ 0 & \tilde{\semtrx{S}}^{s}} 
    \left({\tilde{\semtrx{B}}^{s}}\right)\trn
    \right)
\end{equation}
where $\semtrx{\tilde{S}}^{s}$ either
\begin{itemize}
    \item represents the Schur complement $\semtrx{S}^{s}=\semtrx{K}_{bb}^{s} - \semtrx{K}_{bi}^{s} {\semtrx{K}_{ii}^{s}}^{-1} \semtrx{K}_{ib}^{s}$ if the optimal Dirichlet preconditioner is used, or
    \item $\tilde{\semtrx{S}}^{s} = \semtrx{K}_{bb}$ or $\tilde{\semtrx{S}}^{s} = \text{diag} \, \semtrx{K}_{bb}$ if a computationally cheaper approximations of $\semtrx{S}^{s}$ is used, denoted as a lumped or super-lumped preconditioner, respectively,
\end{itemize}
where $\text{diag}\,\bullet$ extracts only the diagonal part of $\bullet$.
Note that in FETI-DP the unknown index sets $i$ and $b$ are subsets of $r$ only, since the unknowns in set $c$ are handled directly.

In the simplest setting, one can assume $\semtrx{\tilde{B}}^{s} = \semtrx{B}^{s}$; however, $\semtrx{\tilde{B}}^{s}$ can be arbitrarily scaled provided that~\cite{bib:sfeti}
\begin{equation}
    \sum_{s=1}^{N_{s}} \left( \semtrx{B}^{s} (\tilde{\semtrx{B}}^{s})\trn \right)\semtrx{B}^{j} = \semtrx{B}^{j} \,, \quad \forall 1 \leq j \leq N_{s} \,.
\end{equation}
Based on a mechanical reasoning, Rixen and Farhat~\cite{rixen_simple_1999} proposed $k$-scaling, which compensates for different stiffness coefficients across subdomain boundaries. In their approach, $\tilde{\semtrx{B}}^{s} = \semtrx{W}^{s} \semtrx{B}^{s}$ with the entries of the diagonal matrix $\semtrx{W}^{s}$ obtained as
\begin{equation}
	W_{ii}^{s} = \frac{  K_{ii}^{r} }{ K_{ii}^{s} + K_{ii}^{r} } \,,
	\label{eq:k-scaling}
\end{equation}
where $K_{ii}^{s}$ denotes the component of the stiffness matrix pertinent to the $i$th row of $\semtrx{B}^{s}$ and, similarly, $K_{ii}^{r}$ is the corresponding coefficient of the stiffness matrix of the adjacent $\domain^{r}$. In the case of multi-domain constraints (which typically happen for corner nodes in T-FETI), the authors of~\cite{rixen_simple_1999} advocate for the use of pair-wise constraints despite the introduced redundancies. Consequently, only the denominator in \Eref{eq:k-scaling} changes such that it sums the stiffness contributions from all related domains; see~\cite[Eq. 68]{rixen_simple_1999}. 
Also note that \Eref{eq:k-scaling} naturally falls back to multiplicity scaling in the case of domains composed of the same homogeneous material.

\subsection{Full orthogonalization}

For practical applications, $\semtrx{F}$-orthogonalization of the current (projected) preconditioned search direction $\sevek{w}$, which appears in the classical Preconditioned Conjugate Gradient method used to solve problems (\ref{eq:tfeti:FGG0=lambdaalfa}) or (\ref{eq:fetidp:interface}), with respect to the search directions $\sevek{w}$ from the previous iterations is often recommended, e.g.~\cite{bib:sfeti}.
Consequently, this modification necessitates storing the previous directions and their $\semtrx{F}$-norms.
However, the faster convergence caused by this modification usually compensate for both the increased memory requirements and the orthogonalization-related calculations.

\subsection{Rank-Revealing Simultaneous FETI}

Finally, the last enhancement investigated in this work, is the simultaneous variant of both T-FETI and FETI-DP~\cite{bib:sfeti}.
This modification was particularly developed to handle heterogeneous problems by exploiting the additive structure of the local preconditioners. Instead of considering only one preconditioned search direction $\sevek{w}$ in each iteration, it defines an individual search direction for each subdomain. Note that the original search direction can be restored with $\sevek{w} = \semtrx{W} \, \sevek{1}$, where $\sevek{1} = \mtrx{1, 1, \dots 1}\trn \in \setRn{N_s}$ and $\semtrx{W}$ collecting these individual directions as its columns. 

As a result, each iteration $i$ requires solving a system of $N_{s} \times N_{s}$ equations related to minimization and conjugation steps; however, the matrix $\semtrx{\Delta}_{i}$ as defined in Algorithm~\ref{alg:SFETI} may be ill-conditioned, e.g. due to a linear dependence of the search directions, and its pseudo-inverse might need to be constructed. Here, we employ the rank revealing Cholesky factorization~\cite{bib:CholeskyFacotrizationwithPivoting} with permutation matrix $\semtrx{N}$ to extract only the linearly independent search directions, which are subsequently $\semtrx{F}$-orthonormalized; see lines 9--12 in Algorithm~\ref{alg:SFETI}, which summarizes the strategy for T-FETI.

With this modification, we also automatically employ the full orthogonalization described in the previous subsection; lines 18--21 in Algorithm~\ref{alg:SFETI}. Note that in Algorithm~\ref{alg:SFETI}, $\semtrx{P}$ is the projection matrix $\semtrx{P} = \semtrx{I} - \semtrx{G}\trn \left(\semtrx{G}\semtrx{G}\trn\right)^{-1} \semtrx{G}$ enforcing the increments of $\sevek{\lambda}$ to satisfy $\semtrx{G}\,\sevek{\lambda}=\sevek{e}$.

\begin{algorithm} 
\caption{Rank-Revealing Simultaneous FETI~\cite{bib:sfeti}}
\label{alg:SFETI}
\DontPrintSemicolon
	$\sevek{r}_{0} = \semtrx{P}\trn (\sevek{d} - \semtrx{F} \sevek{\lambda}_{0})$ \;
	$\semtrx{Z_{0}} = \mtrx{ \dots, \tilde{\semtrx{B}}^{s} \, \semtrx{\tilde{S}}^{s} ({\tilde{\semtrx{B}}}^{s})\trn \sevek{r}_{0} ,\dots}, \forall s \in \left\{1, \dots, N_{s} \right\}$\; 
	$\semtrx{W_{0}} = \semtrx{P}\,\semtrx{Z}_{0}$\,\; 
	%
	%
	$\tilde{\sevek{\lambda}}_{0} = \sevek{0}$\,\; 
	$i \leftarrow 0$\,\; 
	\While{$\sqrt{\sevek{r}_{i}\trn \, \semtrx{Z}_{i} \,\semtrx{1}}\,>\,\epsilon$}{
		$\semtrx{Q}_{i} = \semtrx{F} \, \semtrx{W}_{i}$\,\; 
		$\semtrx{\Delta}_{i} = \semtrx{Q}_{i}\trn \, \semtrx{W}_{i}$ \; 
		%
		Factorize $\semtrx{N} \, \semtrx{\Delta}_{i} \, \semtrx{N}\trn = \semtrx{L} \, \semtrx{L}\trn$ s.t. $\semtrx{L} = \mtrx{ \tilde{\semtrx{L}} & \semtrx{0} \\ \semtrx{\times} & \semtrx{0} }$ \; 
		%
		$\semtrx{W}_{i} \leftarrow \semtrx{W}_{i} \, \semtrx{N} \trn \, \mtrx{ \tilde{\semtrx{L}}\invtrn \\ \semtrx{0} }$ \;
		$\semtrx{Q}_{i} \leftarrow \semtrx{Q}_{i} \, \semtrx{N} \trn \, \mtrx{ \tilde{\semtrx{L}}\invtrn \\ \semtrx{0} }$ \;
		$\semtrx{\Delta}_{i} \leftarrow \semtrx{I}$ \; 
		$\sevek{\gamma}_{i} = \semtrx{W}_{i}\trn \, \sevek{r}_{i}$ \; 
		$\tilde{\sevek{\lambda}}_{i+1} = \tilde{\semtrx{\lambda}}_{i} + \semtrx{W}_{i} \, \semtrx{\gamma}_{i}$ \; 
		$\sevek{r}_{i+1} = \sevek{r}_{i} - \semtrx{P}\trn \, \semtrx{Q}_{i} \, \semtrx{\gamma}_{i}$ \;
		$\semtrx{Z}_{i+1} = \mtrx{ \dots, \tilde{\semtrx{B}}^{s} \, \tilde{\semtrx{S}}^{s} ({\tilde{\semtrx{B}}}^{s})\trn \sevek{r}_{i+1} ,\dots}$ \;  
		$\semtrx{W}_{i+1} = \semtrx{P} \, \semtrx{Z}_{i+1}$ \; 
		\For{$0 \leq  j  \leq  i$}{
			$\semtrx{\Psi}_{j} = \semtrx{Q}_{j}\trn \, \semtrx{W}_{i+1}$ \; 
			$\semtrx{W}_{i+1} \leftarrow \semtrx{W}_{i+1} - \semtrx{W}_{j} \, \semtrx{\Psi}_{j}$
		} 
		$i \leftarrow i+1$ \; 
	} 
\end{algorithm}

\section{Numerical tests}
\label{sec:Numerical_tests}

\subsection{Academic problems}
\label{sec:academic}

\begin{figure}[h!]
	\centering
	\begin{tabular}{cc}
		\multicolumn{2}{c}{\includegraphics[width=0.9\columnwidth]{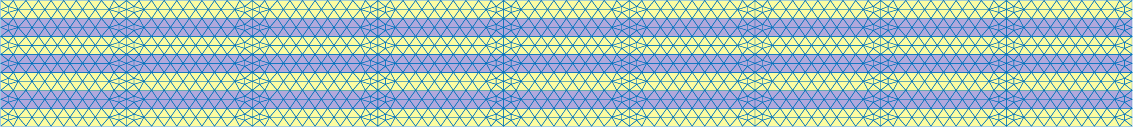}} \\
		\multicolumn{2}{c}{(a)} \\
		\includegraphics[width=0.20\textwidth]{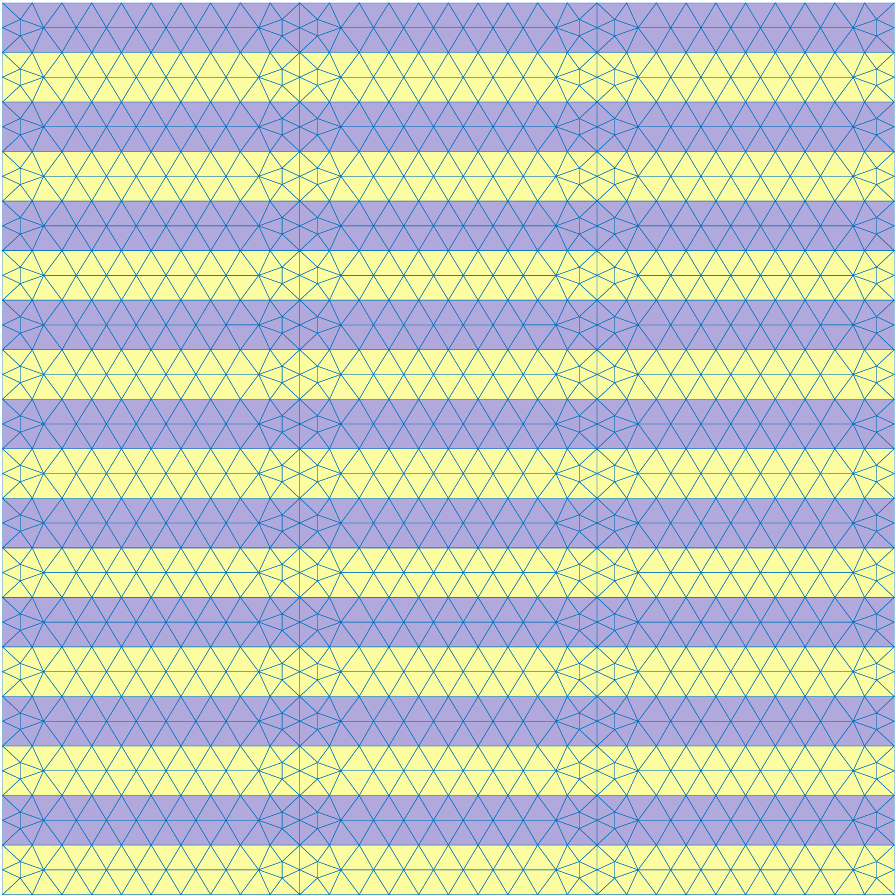} & \includegraphics[width=0.20\textwidth]{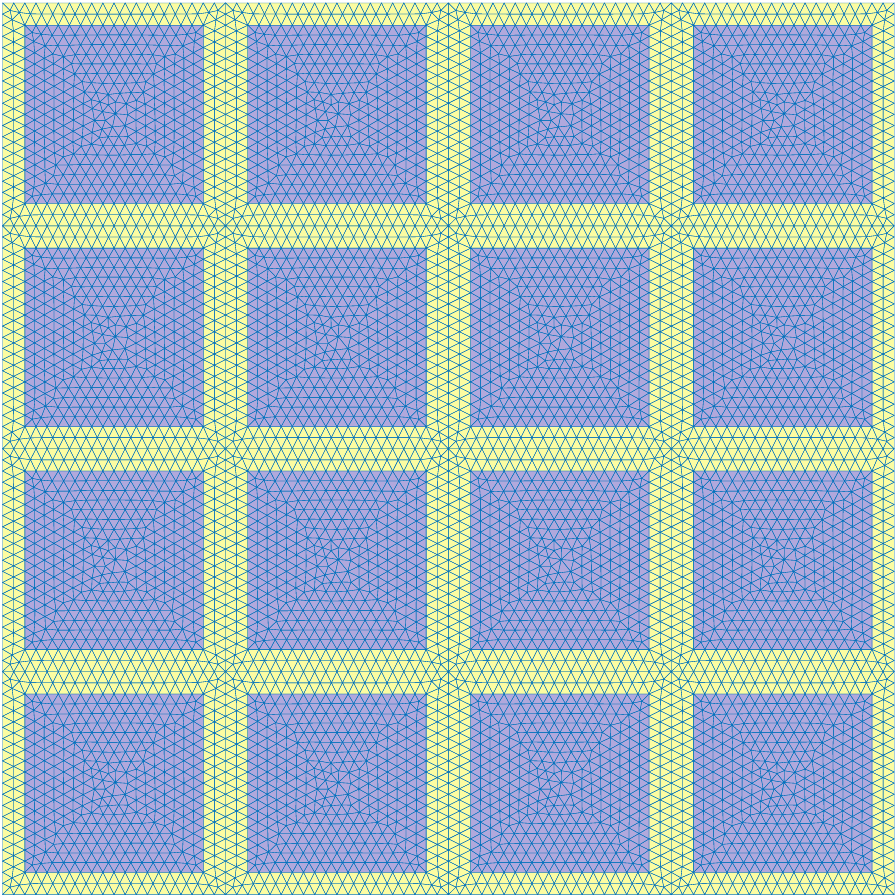} \\
		(b) & (c)
	\end{tabular}
	\caption{Three academic problems taken from~\cite{bib:sfeti}: a)~the laminated beam, b)~the $3 \times 3$ grid of periodic, layered subdomains, and, c) the $4 \times 4$ grid of peridodic subdomains with immersed stiff square inclusion.}
	\label{fig:gosselet}
\end{figure}

We test both approaches on two sets of illustrative examples. The first suite comprises three test problems studied, e.g., in~\cite{bib:sfeti} and shown in~\Fref{fig:gosselet}. 
Unlike~\Fref{fig:gosselet}, however, we keep the regular partitioning and do not investigate the effect of irregularity and bad aspect ratios of the decomposition.
We included these problems primarily to verify our implementation against the convergence reported in~\cite{bib:sfeti} and also to compare the performance of all considered variants in these academic problems against the problems arising in topology optimization.

The first test is a laminated beam consisting of nine repeating square subdomains, each composed of seven alternating layers of stiff (blue) and compliant (green) material. 
The second test problem is a square grid of $3 \times 3$ subdomains, again, with alternating layers. This problem features the corner cross-point and hence different variants of enforcing the continuity of solution is expect to deliver different performance results.
Finally, the last test problem is a $4 \times 4$ grid of subdomains with a stiff inclusion immersed in each subdomain. Unlike the previous two problem, there are no jumps in stiffness coefficients and hence the convergence should be the fastest.
All structures were clamped on the left-hand side and subjected to tangential and tensile normal traction on their right-hand sides.

\begin{figure}[!htb]
	\setlength{\tabcolsep}{0pt}
	\centering
	\begin{tabular}{cc}
		(a) & \includegraphics[width=0.75\linewidth]{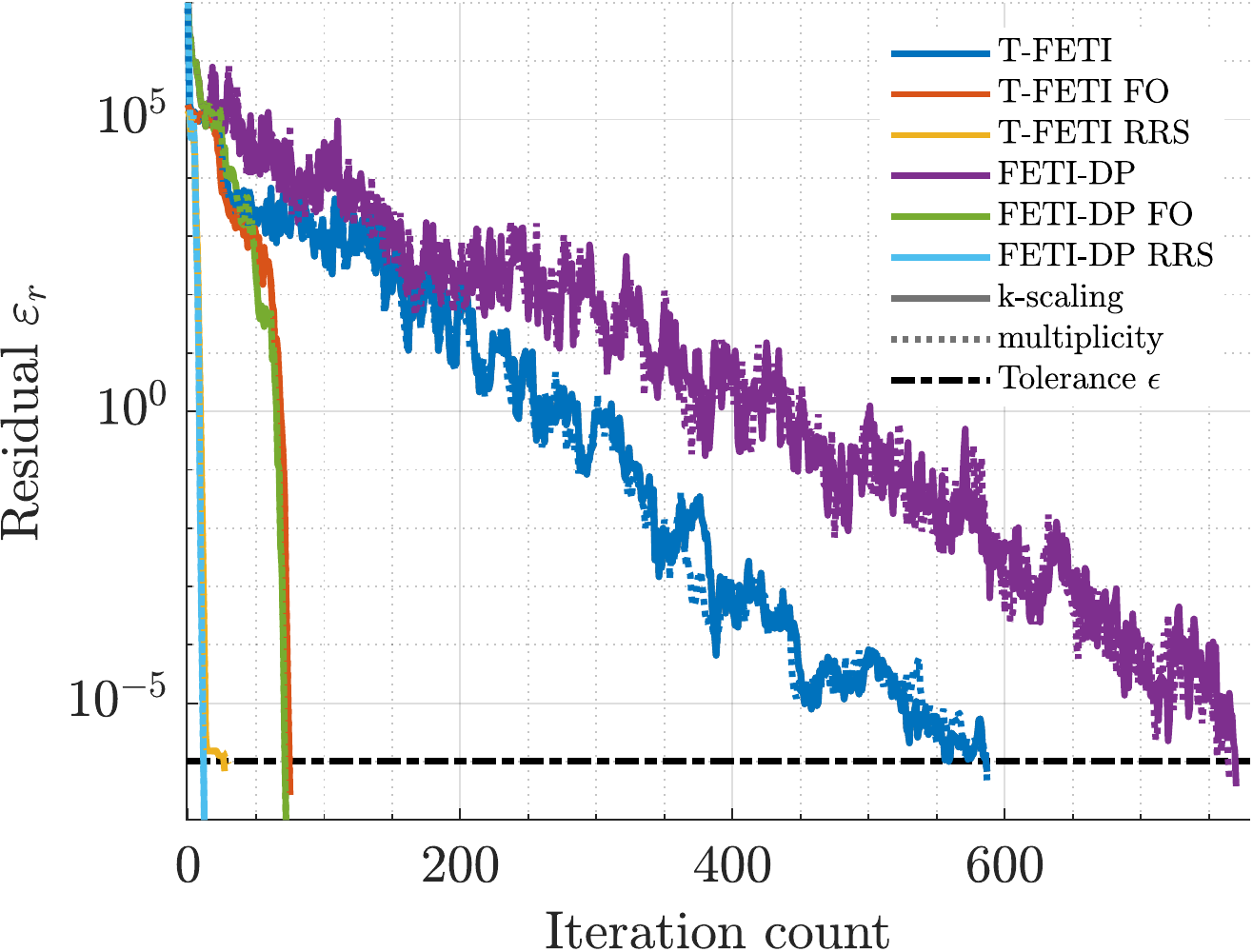} \\
		(b) & \includegraphics[width=0.75\linewidth]{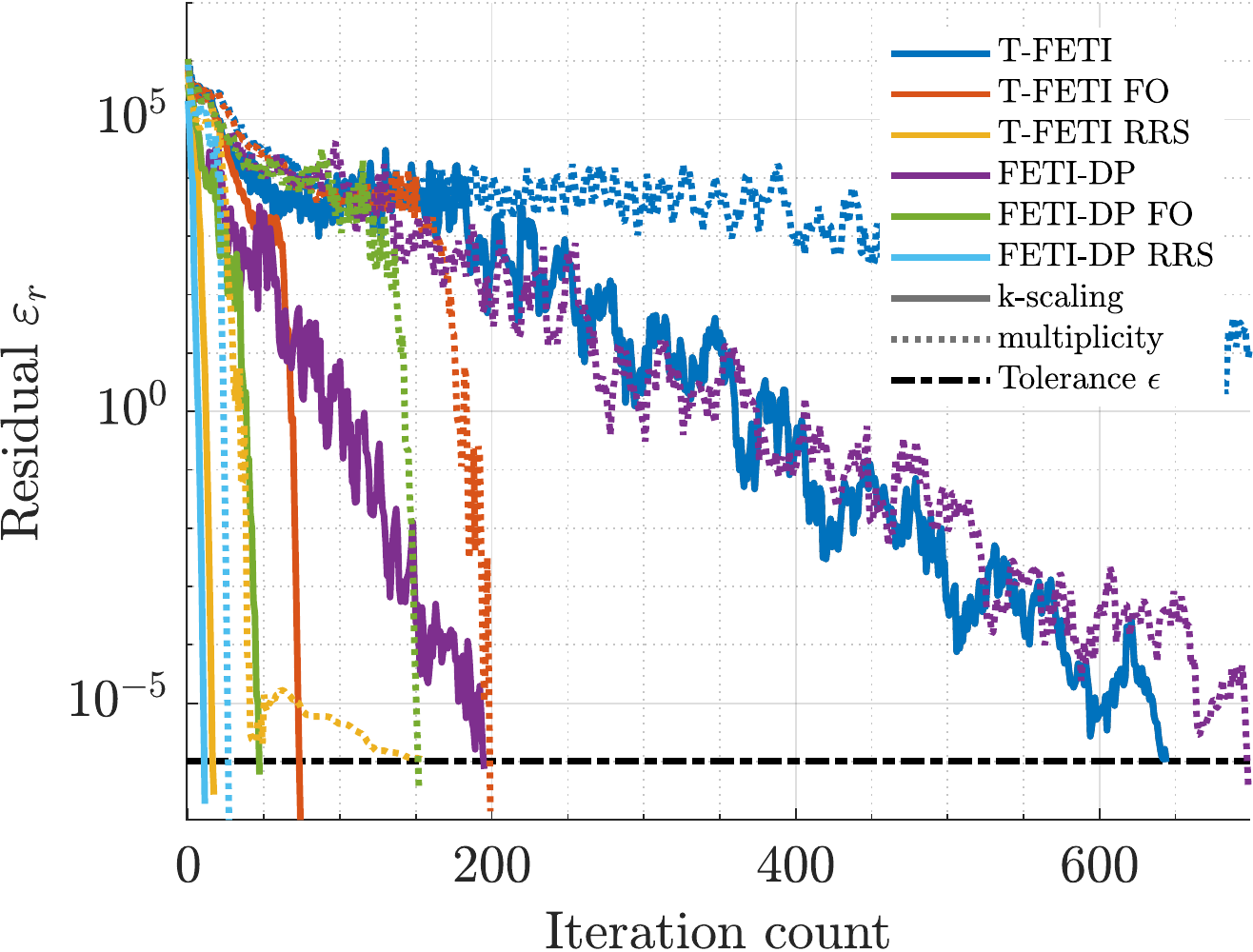} \\
		(c) & \includegraphics[width=0.75\linewidth]{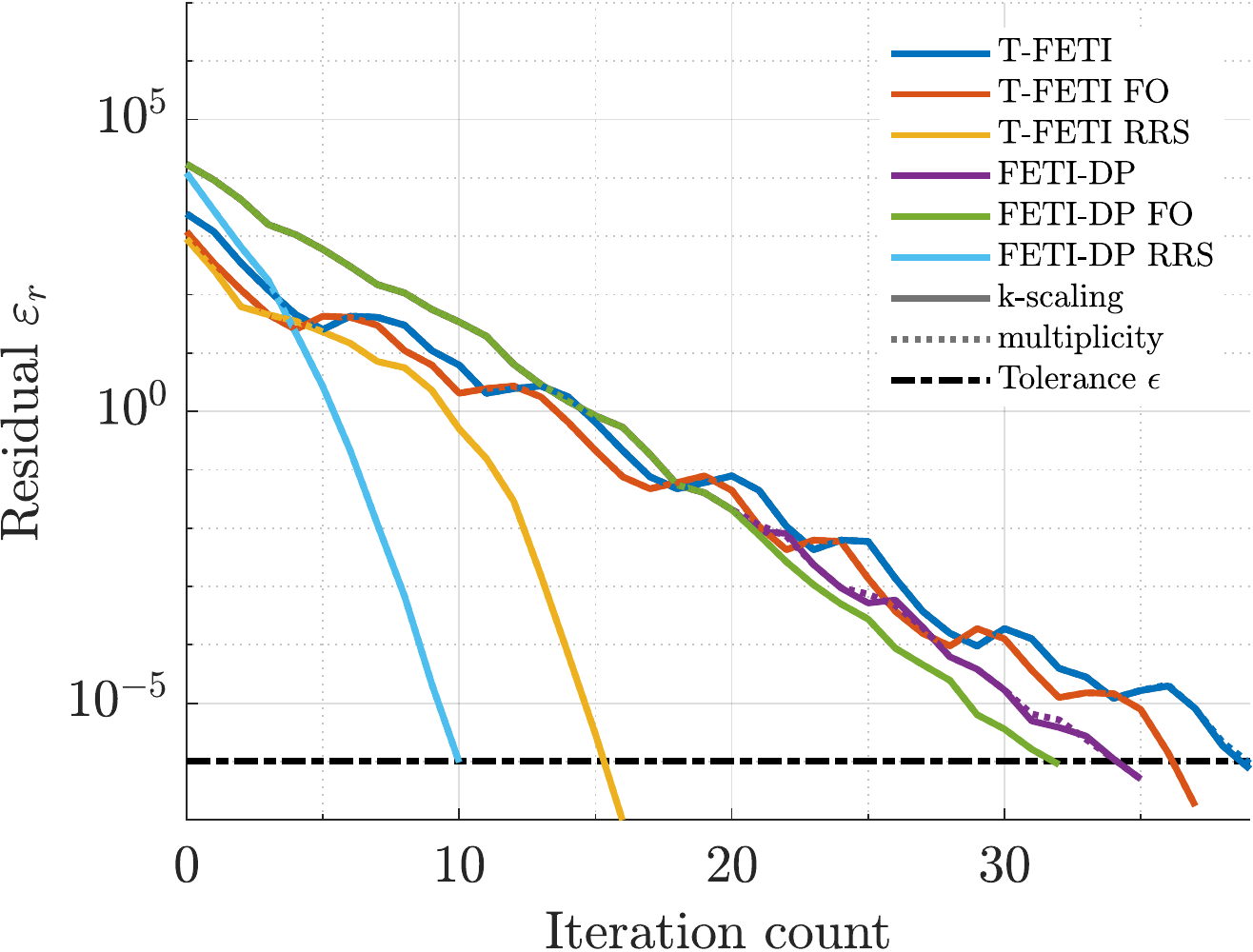}
	\end{tabular}
	\caption{Convergence history of 12 variants of the solution strategies for the three academic problems shown in~\Fref{fig:gosselet}: a)~a laminated beam, b)~a $3 \times 3$ grid of laminated subdomains, and, c) a $4 \times 4$ grid of subdomains with stiff square inclusions.
	FO denotes full orthogonalization enhancement, RRS stands for the simultaneous variant with rank-revealing decomposition. In all graphs, results with the $k$-scaling enhancement are plotted in solid lines while the dotted lines represent the multiplicity scaling. The dash-dotted line mark the desired residual threshold $\epsilon$ from Algorithm~\ref{alg:SFETI}.}
	\label{fig:results_academic}
\end{figure}

The two solution strategies (T-FETI vs. FETI-DP) in combination with the scaling options (stiffness vs. multiplicity) and the option for full orthogonalization or simultaneous search directions yield 12 variants in total. Figure~\ref{fig:results_academic} plots the history of the residual error defined as
\begin{equation}
	\varepsilon_{r} = \sqrt{ \sevek{r}_{i}\trn \sevek{z}_{i} } \,,
	\label{eq:residual}
\end{equation}
where the $\sevek{r}_{i}$ and $\sevek{z}_{i}$ are the projected residual and its preconditioned counterpart from  Algorithm~\ref{alg:SFETI}, respectively.

In accordance with expectations, the third problem with stiff square inclusions was indeed the easiest to solve and the performance of all variants did not vary significantly except for the most advanced simultaneous variants of both T-FETI and FETI-DP which converged in fewer than half of the iterations needed by the other variants.
On the other hand, once the heterogeneity in stiffness coefficients reached subdomain boundaries (the first and the second test problems), the performance of the unmodified variants deteriorated rapidly.
However, the importance of $k$-scaling (solid lines) in contrast to the simple multiplicity scaling (dotted lines) arises only in the second example due to the presence of the cross-points; in the third problem for instance, the $k$-scaling falls back to multiplicity scaling as mentioned in~\Sref{sec:k-scaling}.

\subsection{Topology optimization problems}

\begin{figure}[h!]
	\centering
	\includegraphics[width=0.95\linewidth]{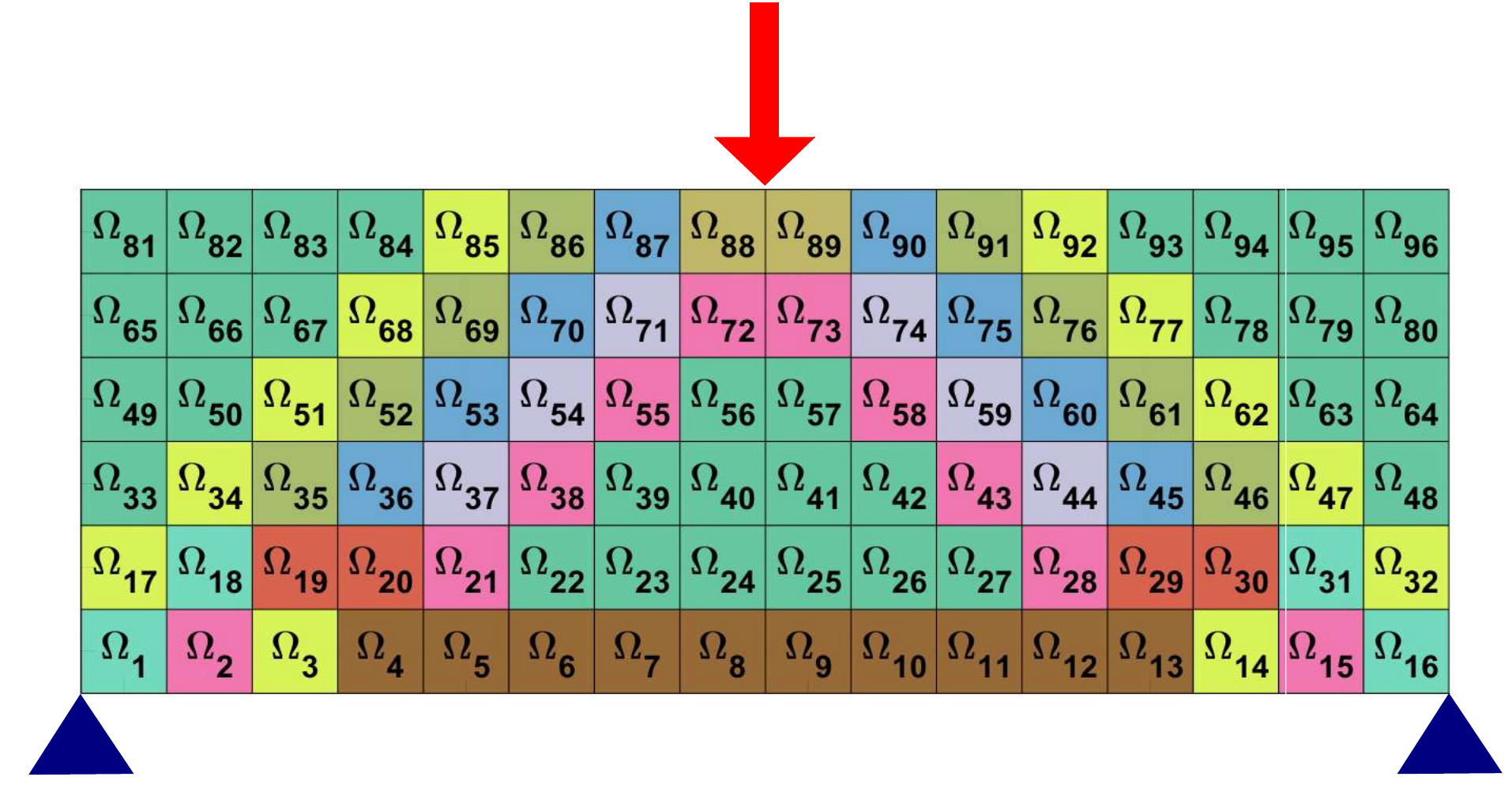} 
	\caption{An illustration of the beam problem arising in modular topology optimization. The distinct colours represent 16 different types of 96 modules, which constitute the beam's subdomains $\Omega^{s}$.}
	\label{fig:ModulesAssemblyPlan}
\end{figure}

The second test suite comprises several snapshots from the modular topology optimization~\cite{bib:TrussStructures} of a beam fixed at both bottom corners and loaded by a unit force acting at the midspan of the top edge; see~\Fref{fig:ModulesAssemblyPlan}. The beam is composed of 96 square modules of 16 types (depicted with distinct colours in~\Fref{fig:ModulesAssemblyPlan}). The modular composition was directly reflected in the beam's partitioning into 96 subdomains.  Each subdomain was discretized with $30 \times 30$ bilinear quadrilateral finite elements, which resulted in 184,512 DOFs in total and over 10,000 DOFs in the interface problem of both T-FETI and FETI-DP.

\begin{figure}[h!]
	\centering
	\setlength{\tabcolsep}{2pt}
	\begin{tabular}{ccc}
	    (a) & \includegraphics[width=0.335\textwidth]{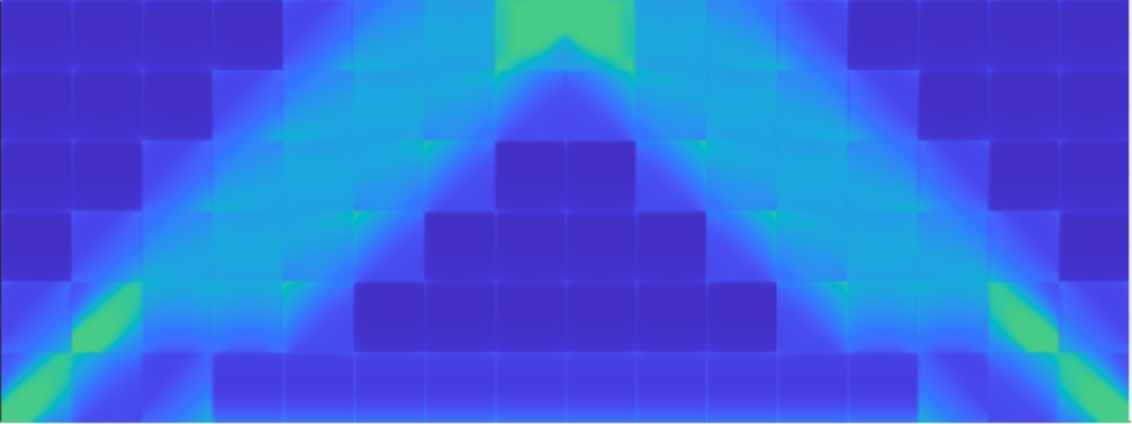} & \multirow[t]{3}{*}[-4.63cm]{\includegraphics[height=6.85cm]{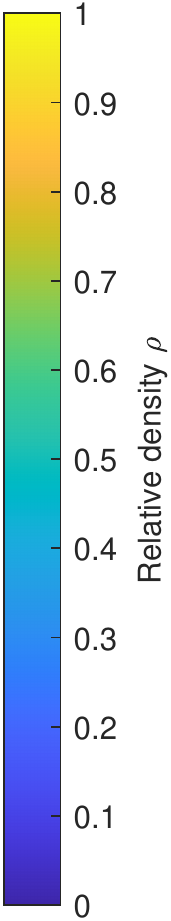}}\\
		(b) & \includegraphics[width=0.335\textwidth]{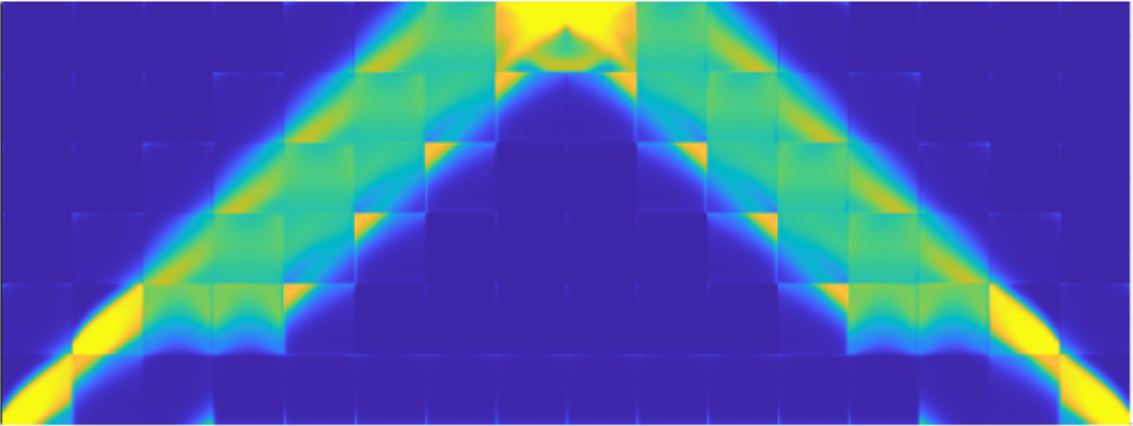} & \\
		(c) & \includegraphics[width=0.335\textwidth]{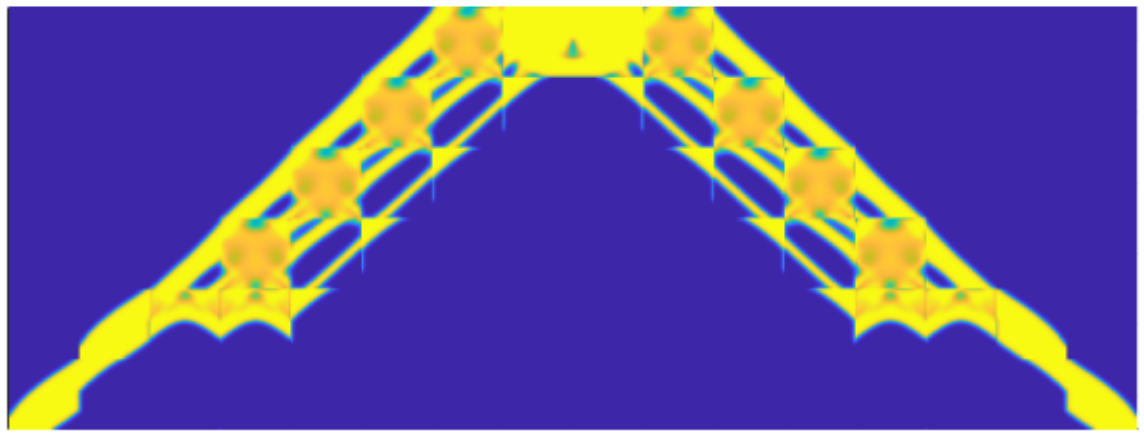}  
	\end{tabular}
	\caption{Three snapshots of modular topology optimization constituting the second suite of test problems: a) $4^{\text{th}}$ iteration, b) $8^{\text{th}}$ iteration, and c) $30^{\text{th}}$ iteration.}
	\label{fig:Steps4830}
\end{figure}

In particular, we tested the described solution strategies on the $4^{\text{th}}$, $8^{\text{th}}$, and $30^{\text{th}}$ iteration snapshots of the topology optimization.
Distribution of the base material, parameterized by the relative density $\rho$ (recall \Eref{eq:SIMPstiffness}), in these snapshots is shown in~\Fref{fig:Steps4830}. As the topology optimization progresses, the initially uniform density gradually concentrates to the most efficient locations, which leads to a highly heterogeneous problem with significant jumps in stiffness coefficients along subdomain boundaries, making the problem difficult to solve with iterative methods.

\begin{figure}[!h]
	\setlength{\tabcolsep}{0pt}
	\centering
	\begin{tabular}{cc}
		(a) & \includegraphics[width=0.75\linewidth]{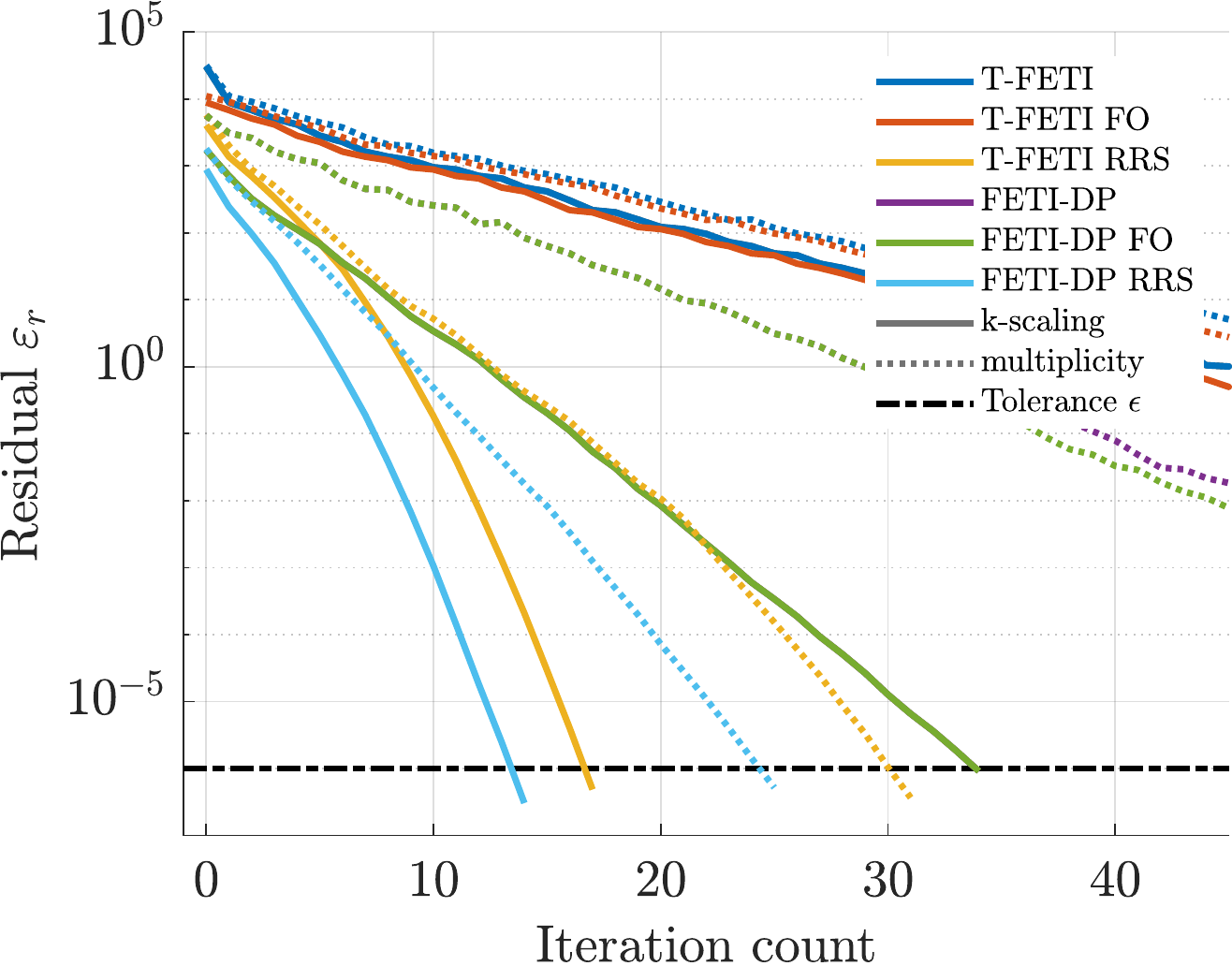} \\
		(b) & \includegraphics[width=0.75\linewidth]{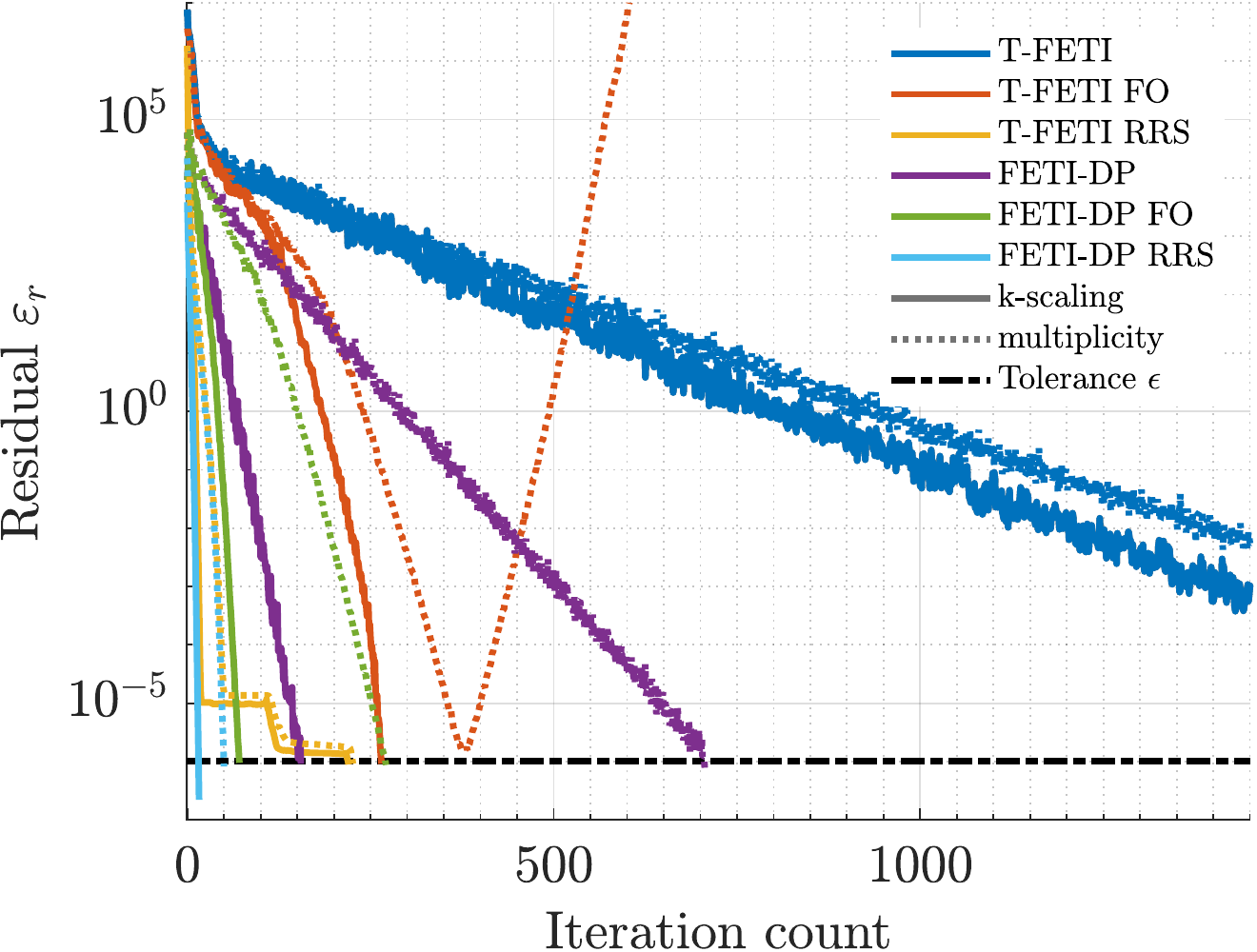} \\
		(c) & \includegraphics[width=0.75\linewidth]{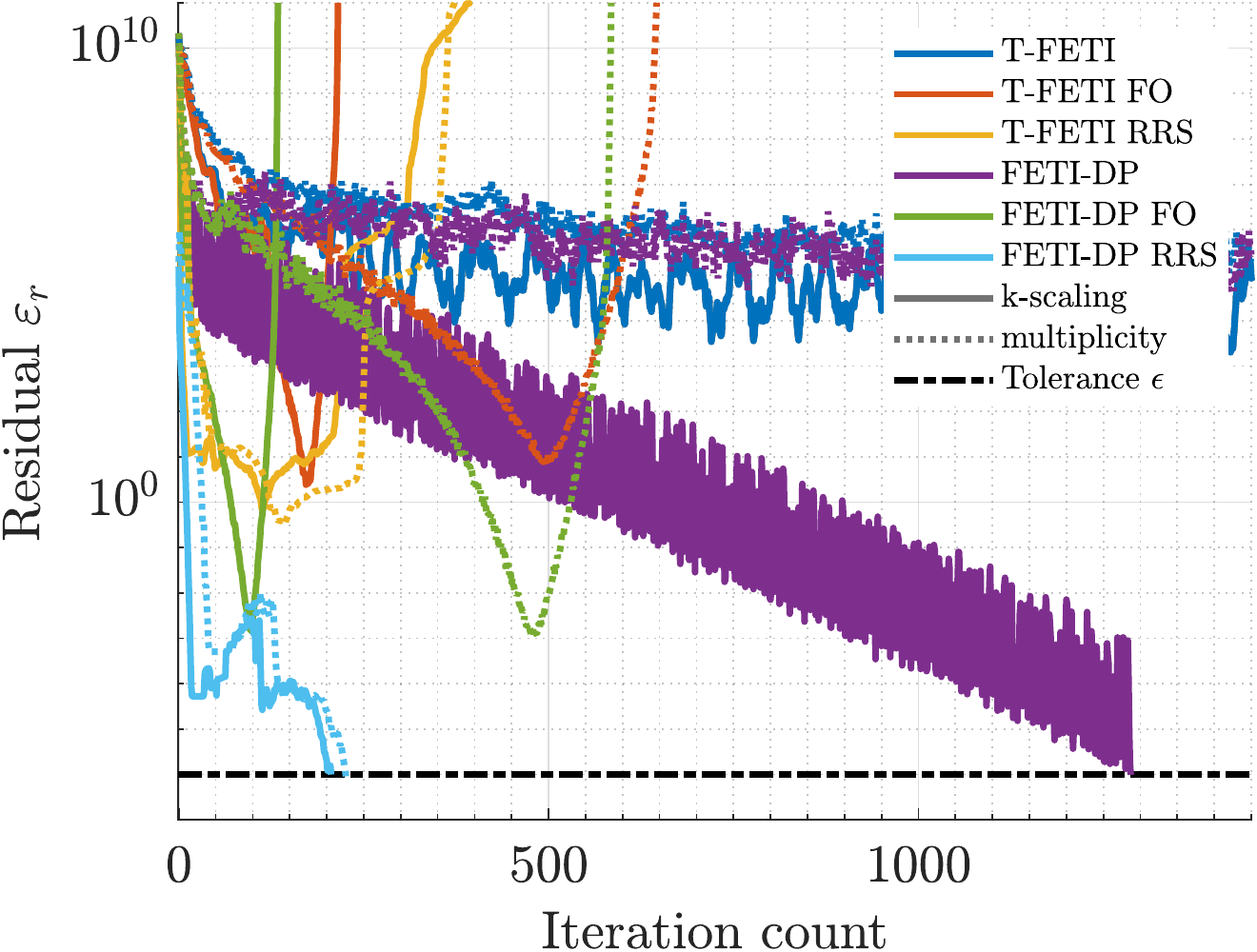}
	\end{tabular}
	\caption{Convergence history of 12 variants of the solution strategies for the three snapshots of modular topology optimization from~\Fref{fig:Steps4830}: a) $4^{\text{th}}$, b) $8^{\text{th}}$, and c) $30^{\text{th}}$ iteration of the topology optimization.
	Similarly to~\Fref{fig:results_academic}, FO denotes full orthogonalization enhancement, RRS stands for the simultaneous variant with rank-revealing decomposition. In all graphs, results with the $k$-scaling enhancement are plotted in solid lines while the dotted lines represent the multiplicity scaling. The dash-dotted line mark the desired residual threshold $\epsilon$ from Algorithm~\ref{alg:SFETI}.}
	\label{fig:results_topopt}
\end{figure}

Similarly to Section~\ref{sec:academic}, Figure~\ref{fig:results_topopt} summarizes the convergence history of all 12 variants for the three topology optimization snapshots from~\Fref{fig:Steps4830}. 
The later stages of the topology optimization in particular render challenging problems for T-FETI and FETI-DP solvers; while in the first snapshot, all methods except for T-FETI with full orthogonalization successfully converged, in the last snapshot, only the simultaneous FETI-DP with rank-revealing factorization succeeded.

\section{Discussion}

The numerical tests clearly demonstrate that the linear systems of equations arising in modular topology optimization are indeed challenging for domain-decomposition based iterative methods. The main culprit is the potentially unfavorable partitioning into subdomains, which leads to coefficients jumps across subdomain boundaries, and most importantly the high contrast in stiffness coefficients. These effects are partly present in the academic test problems, however, they are significantly more pronounced in the topology optimization problems up to the point at which most of the considered solution variants break or do not converge in a reasonable time.

The only variant successful in all tests was the simultaneous FETI-DP with rank-revealing Cholesky decomposition. With incorporated $k$-scaling, this variant was also the fastest to converge, even though the convergence rate of T-FETI with rank-revealing was usually comparable when it converged.

On the other side of the convergence spectrum were the original variants of T-FETI and FETI-DP. However, while featuring the slowest convergence, those methods proved to be relatively robust, unlike the modifications with full orthogonalization, which reduced the needed iterations only for mildly heterogeneous problems and broke completely for the final test problem.
We conjecture that the rapid increase in the monitored residual in~\Fref{fig:results_topopt}c is caused by the loss of numerical accuracy due to round-off errors during the orthonormalization.

\section{Summary}

In this work, we investigated twelve variants of dual and dual-primal domain decomposition methods with particular emphasis on applications to highly heterogeneous problems with predefined partitioning which arise in modular topology optimization problems~\cite{bib:TrussStructures,tyburec_modular_2022}. 
We considered Total-FETI~\cite{bib:tfeti} and FETI-DP~\cite{bib:feti-dp} as the base methods, for which we tested two scaling approaches ($k$-scaling and multiplicity scaling), full orthogonalization, and simultaneous search directions with rank-revealing decomposition.
We tested the variants on two suites of two-dimensional elasticity problems: first one composed of three academic problems, the second suite comprising snapshots of modular topology optimization.
All variants were compared in terms of the number of iterations needed to reach a desired residual $\epsilon$. We are aware that such a comparison does not paint the whole picture and comparing the total wall-clock time would be optimal; however, such a comparison would be heavily biased by the implementation and available hardware.

Based on our results, we conclude that the $k$-scaling should be a default setting when dealing with non-homogeneous problems thanks to its significant impact on the convergence rate and negligible computational overhead.
Our results also show that the original formulations of both T-FETI and FETI-DP exhibit very slow convergence in the presence of severe heterogeneity in stiffness coefficients and hence performance enhancements are needed. The simultaneous search directions with the rank-revealing Cholesky decomposition reduced the number of needed iterations the most and should be therefore preferred to full orthogonalization alone.
Finally, FETI-DP proved to be more robust than T-FETI as it was the only method capable of reaching the desired residual in all test problems.

Following these conclusions, our current research interest lies in (i) an adaptive choice of the nodes whose compatibility should be enforced directly in primal variables (i.e., not necessarily choosing only the corner nodes for the primal part of FETI-DP) and (ii) identification of boundary modes responsible for the bad convergence and eliminating from the iterative process in the spirit of~\cite{feti_geneo}.

\begin{acknowledgements}
	The authors would like to thank Prof. Jaroslav Kruis and Dr. Petr Mayer, both from the Czech Technical University in Prague, for invaluable discussions and suggestions.
	TM, MD, and JZ also gratefully acknowledge support by the Czech Science Foundation, project No. 19-26143X.
\end{acknowledgements}

\bibliographystyle{actapoly}
\bibliography{biblio}

\end{document}